\documentclass[%
 reprint,
 amsmath,amssymb,
 aps,
 nofootinbib,
]{revtex4}
\usepackage{dsfont}
\usepackage{graphicx}
\usepackage{dcolumn}
\usepackage{bm}
\usepackage{hyperref}

\newcommand{\fsc}{\alpha_\mathrm{em}}

\def\slash#1{\setbox0=\hbox{$#1$}  
   \dimen0=\wd0     
   \setbox1=\hbox{/} \dimen1=\wd1  
   \ifdim\dimen0>\dimen1   
      \rlap{\hbox to \dimen0{\hfil/\hfil}} 
      #1     
   \else     
      \rlap{\hbox to \dimen1{\hfil$#1$\hfil}} 
      /      
   \fi}      %

\newcommand{\sumint}{\sum_{X}\hspace{-0.5cm}\int}

\begin{document}

\preprint{arXiv:2405.04232 [hep-ph]}

\title{Probing the polarized photon content of the proton in $ep$ collisions at the EIC}

\author{Daniel Rein}
\email{da.rein@student.uni-tuebingen.de}
\author{Marc Schlegel}
 \email{marc.schlegel@uni-tuebingen.de}
 \author{Werner Vogelsang}
 \email{werner.vogelsang@uni-tuebingen.de}
\affiliation{
 Institute for Theoretical Physics, University of T\"ubingen, Auf der Morgenstelle 14, D-72076 T\"ubingen, Germany
}

\date{\today}

\begin{abstract}
We study the single-inclusive production of prompt photons in electron proton collisions, $ep \to \gamma X $, for kinematics relevant at the Electron-Ion Collider (EIC). 
We perform a perturbative calculation of the differential cross section to next-to-leading order in QCD and to lowest order in QED. We consider unpolarized 
collisions as well as scattering of longitudinally polarized incident electrons and protons. We show that the cross sections are sensitive to the 
parton distribution functions of photons inside the proton, which we find to generate the dominant contributions in certain kinematical regions at the EIC. We also
investigate the effects of photon isolation on the unpolarized and polarized cross sections.
\end{abstract}

\maketitle

\section{\label{sec:Intro}Introduction}

Production of photons with large transverse momentum in hadronic scattering serves as an important probe
of the partonic structure of hadrons, especially of their gluon distributions. Compared
to hadronic final states, photons offer the advantage that a substantial  -- and often strongly dominant -- 
part of the production mechanism comes from ``point-like'' contributions for which the produced photon couples 
directly to a partonic hard-scattering process. In the present paper, we study 
photon production in a different and somewhat simpler setting, in electron proton collisions. Previous studies in this area
have considered the process $ep\to e\gamma X$, with detected final-state electron.
Here we will address its truly {\it single-inclusive} counterpart, $ep\to\gamma X$, for which the scattered 
electron is not explicitly observed. 
Our motivation for investigating $ep\to\gamma X$ is twofold. First, as we will show, the process may offer access to the
proton's {\it photon} parton distribution function (PDF). Second, the process is also interesting from a theoretical
point of view, especially when the proton is transversely polarized. Although we do not consider this particular case here,
our calculation provides an important basis for future dedicated studies. We will now describe our two 
motivations in a little more detail. 

There is a long history of interest in the proton's photon PDF, starting
from discussions of its possible relevance for new-physics searches at hadron 
colliders~\cite{Drees:1988pp,Ohnemus:1993qw,Drees:1994zx,Ohnemus:1994xf,Gluck:1994vy}. 
As part of the QED corrections to proton structure, the distribution soon became part of the global analysis of 
PDFs~\cite{Martin:2004dh,Ball:2013hta,Schmidt:2014aba}. Important progress was made by 
Refs.~\cite{Manohar:2016nzj,Manohar:2017eqh} which showed that the photon PDF may in fact be computed from first principles,
although the result relies on elastic contributions as well as on understanding of the proton structure functions down to low scales.
The formalism of~\cite{Manohar:2016nzj,Manohar:2017eqh} has become the paradigm for modern 
studies of the photon content of the proton when incorporated in global analyses of proton 
structure~\cite{Bertone:2017bme,Harland-Lang:2019pla,Cridge:2021pxm}.

The usefulness of electron-proton scattering in obtaining information on the proton's photon PDF 
has been established in numerous studies, primarily in the context of the HERA collider~\cite{Bawa:1988qs,Bawa:1990qx,Aurenche:1992sb,Courau:1992ht,Blumlein:1993ef,Gordon:1994sm,Gordon:1994km,DeRujula:1998yq,Hoyer:2000mb,Fontannaz:2001ek,Gluck:2002fi,Gluck:2002cm,Mukherjee:2003yh,Mukherjee:2004dw,Mukherjee:2004hs,Lendermann:2003rq,H1:2004jab,ZEUS:2009onz,Gehrmann-DeRidder:2006zbx,Gehrmann-DeRidder:2006lpc}. The photon PDF is also central to studies of QED radiative corrections to the DIS and semi-inclusive DIS cross 
sections~\cite{Bardin:1996ch,Liu:2021jfp,Akushevich:2023jjc}. However, to our knowledge, single-inclusive production of photons has not been considered so far in the literature, 
and it is useful to explore its potential ability to put further constraints on the photon PDF, especially so in 
view of the future Electron-Ion Collider (EIC) now under construction~\cite{Accardi:2012qut,AbdulKhalek:2021gbh,AbdulKhalek:2022hcn,Burkert:2022hjz,Amoroso:2022eow,Hentschinski:2022xnd,Abir:2023fpo}. Here our focus will also 
be on the polarized (helicity) photon PDF, whose presence in $ep\to e\gamma X$ has been discussed in 
Refs.~\cite{Gluck:2002fi,Gluck:2002cm,Mukherjee:2003yh,Mukherjee:2004dw,Mukherjee:2004hs,Blumlein:2004bs}. Most of the previous
studies did not take into account the contributions arising from production of photons in jet fragmentation, which
we will develop here to full next-to-leading order (NLO). 

Single-inclusive processes have also been of more general interest in the context of achieving a better understanding 
of single-transverse spin asymmetries (TSSAs)  in hadronic scattering. The theoretical description of TSSAs in reactions
of the type $p^\uparrow p\to h X$ in terms of collinear factorization has turned out to be remarkably complex~\cite{Qiu:1991pp,Qiu:1991wg,Qiu:1998ia,Kouvaris:2006zy,Kanazawa:2010au,Kang:2011hk,Kanazawa:2014dca,Benic:2024fvk}, 
even at the lowest order (LO) of QCD perturbation theory. The main reason for this is that the asymmetry
is power-suppressed in the hard scale of the reaction, the transverse momentum of the produced hadron, and 
hence involves twist-3 hadronic matrix elements and correspondingly hard-scattering diagrams with three partons connecting 
to these matrix elements. As the leading-power unpolarized cross section shows, NLO QCD corrections
are vital for a successful phenomenology of $pp\to h X$~\cite{Jager:2002xm}. For the TSSA, the computation
of NLO corrections is extremely complicated and, although essential for phenomenology, is currently not on the horizon. 
As a first step in this direction, it is useful to consider reactions with similar kinematics, but less complexity. 
One possibility is to go from purely hadronic scattering to $ep$ reactions, essentially replacing one initial 
proton by the simpler electron. References~\cite{Kang:2011jw,Gamberg:2014eia,Kanazawa:2015ajw,Anselmino:1999gd,Anselmino:2009pn,DAlesio:2017nrd,Fitzgibbons:2024zsh}
proposed the process $ep \to \pi X$ in this context, which -- as it turns out -- by itself could offer promising
opportunities to gain new information on twist-3 correlation functions in the proton at the EIC. 
NLO corrections to the spin-averaged cross section for $ep \to \pi X$ were presented in~\cite{Hinderer:2015hra}. 

The process $ep\to\gamma X$ which we consider in the present paper, is an equally valuable (and in some
ways simpler) testing ground for QCD calculations. One may consider it a less complex version of the analogous 
single-inclusive photon production in proton collisions, $pp\to\gamma X$. Historically, the latter reaction with transverse polarization,
$p^\uparrow p\to\gamma X$, played an important role in finding explanations for transverse spin asymmetries~\cite{Qiu:1991pp,Qiu:1991wg}. 
We stress again that the process we investigate differs from the semi-inclusive production of photons in electron-proton collisions,
$ep\to e\gamma X$, for which the final-state electron is detected. The TSSA for the latter has been proposed in Ref.~\cite{Albaltan:2019cyc} 
as an observable well-suited to obtain insights into higher-twist quark-gluon-quark correlation functions. 

In order to set the stage for future work on the TSSA for single-inclusive photon production $ep\to\gamma X$, we present
in this paper a full NLO calculation of the unpolarized and the longitudinally polarized (helicity dependent) cross sections for this reaction, 
which are both of leading twist. Focusing on inelastic contributions, we carefully discuss the various production channels. 
We present phenomenological studies for the EIC, identifying kinematical regions that are favorable with respect to obtaining information 
on the photonic distributions and fragmentation functions. We hope that establishing the basic NLO leading-twist theoretical
framework for $ep\to\gamma X$ will prove useful for future NLO studies of the TSSA for the process. 

Our paper is organized as follows. In section \ref{sec:UCS} we first discuss the various LO and NLO contributions to the differential unpolarized cross section for 
$e p\to \gamma X$ and introduce the contributing PDFs and fragmentation functions, especially the photonic ones. We then 
present the analytic calculation of the full unpolarized cross section. Section~\ref{sec:LCS} addresses the corresponding spin-dependent
cross section $\vec{e} \vec{p} \to \gamma X$. Based on these results, in section \ref{sec:Numerics} we present our numerical predictions
 for the cross sections and longitudinal double spin asymmetries at the EIC. We conclude in section \ref{sec:Conc}.

\section{\label{sec:UCS}The unpolarized cross section}

\subsection{LO and NLO contributions\label{sub:LOandNLO}}

The process we consider is the single-inclusive production of prompt photons in electron-proton collisions, $e(l)+p(P)\to \gamma(P_\gamma)+X$, where we have denoted the relevant four-momenta. In contrast to the reactions usually considered in electron-proton scattering, $X$ denotes an 
unobserved multi-particle final state that consists not only of the hadronic remnants of the initial proton, but also of the final-state electron. 
At large c.m. energies $\sqrt{s}$ and reasonably large transverse momenta of the detected photons we may assume that collinear factorization may be used to 
write the cross section in terms of short-distance partonic hard-scattering cross sections and long-distance hadronic matrix elements. 

\begin{figure*}
\includegraphics[width=0.7\textwidth]{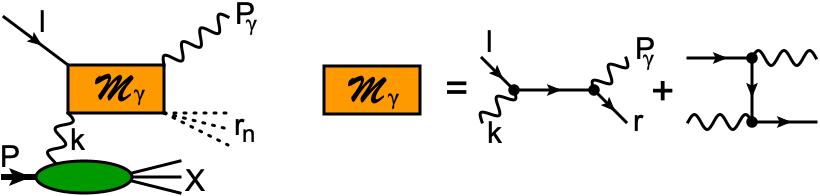}\\[5mm]
\includegraphics[width=0.6\textwidth]{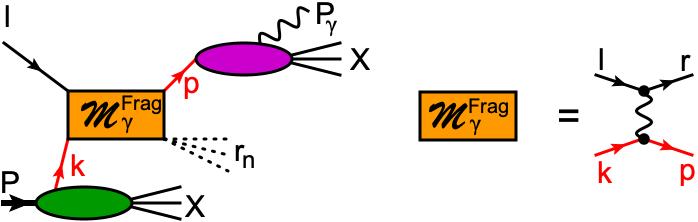}
\caption{\label{fig:LO} LO contributions to $e(l)+p(P)\to \gamma(P_\gamma)+X$. 
We show sketches of factorized amplitudes where photons play the role of a parton in the proton
(upper panel), 
and where photons are produced in a quark fragmentation process (lower panel).}
\end{figure*}

We will work to the lowest order in the QED coupling $\alpha_{\mathrm{em}}$, but will derive the cross section for the process to NLO in QCD. 
The counting of perturbative orders in the strong coupling $\alpha_s$ for the process is somewhat special, but familiar from other hadronic processes with 
identified photons. At LO, the process can proceed in two ways: (i) a photon plays the role of a parton inside the proton and participates in a QED Compton reaction; (ii) a QED electron-quark scattering process is followed by fragmentation of
the quark to a photon. The two types of contributions are depicted in Fig.~\ref{fig:LO}. Both the photon PDF of the proton
and the quark-to-photon fragmentation functions are of perturbative order ${\cal O}(\alpha_{\mathrm{em}}/\alpha_s)$, owing to an explicit logarithm
arising in the calculation of higher-order diagrams. Thus, when combined with the overall factor $\alpha_{\mathrm{em}}^2$ 
arising from the Compton or the $eq$ scattering process, respectively, the lowest order for the process $e(l)+p(P)\to \gamma(P_\gamma)+X$ becomes $\alpha^3_{\mathrm{em}}/\alpha_s$. 
We note that the photon PDF also includes an elastic part for which the proton remains intact and that gives rise to an 
exclusive final state~\cite{Kniehl:1990iv,DeRujula:1998yq,Mukherjee:2003yh,Manohar:2016nzj,Manohar:2017eqh}.
Such elastic contributions are thus included in our analysis. 

\begin{figure*}
\includegraphics[width=0.8\textwidth]{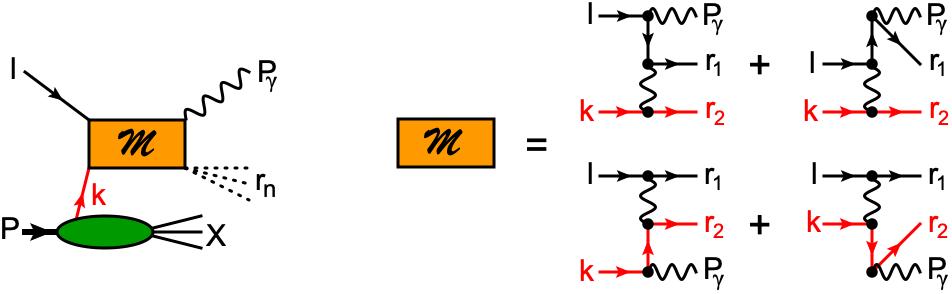}
\caption{\label{fig:PM} NLO contributions to $e(l)+p(P)\to \gamma(P_\gamma)+X$. The hard scattering amplitude consists of four diagrams 
where the detected photon is radiated off the electron line or off the quark line.}
\end{figure*}

NLO contributions arise at order $\alpha_s\times \alpha^3_{\mathrm{em}}/\alpha_s=\alpha^3_{\mathrm{em}}$. They again come in two
classes. The first class consists of all $2\to 3$ diagrams for $e q\to \gamma e q$, as shown in Fig.~\ref{fig:PM}.  
In the context of semi-inclusive photon production, $ep\to e\gamma X$, contributions associated with these diagrams are often 
referred to as ``virtual Compton scattering'' contributions in the 
literature~\cite{Mukherjee:2003yh,Mukherjee:2004dw,Mukherjee:2004hs}. For the single-inclusive
process we consider, the outgoing electron and quark are not observed, so that their momenta need to be integrated over their full phase
space. The corresponding calculations will be presented below. Treating quarks as massless, the diagrams will exhibit collinear 
singularities when the incoming quark splits into a quark+photon pair, or when the outgoing quark radiates the observed photon
collinearly. These singularities are absorbed into the photon PDF of the proton and the quark-to-photon fragmentation function, respectively,
giving rise to the ``inhomogeneous'' part of the evolution equations for these functions in the LO contribution.  

\begin{figure*}
\includegraphics[width=0.7\textwidth]{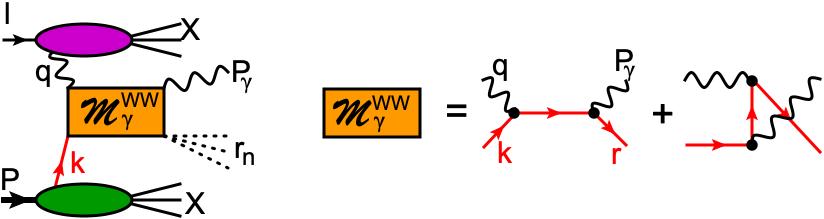}\\[5mm]
\includegraphics[width=0.6\textwidth]{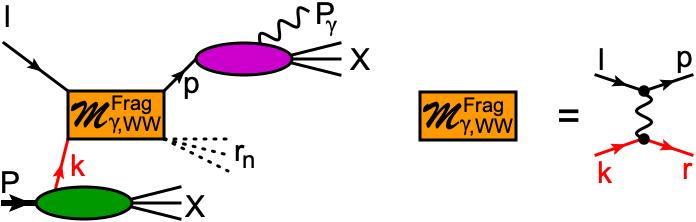}
\caption{\label{fig:PDFWW} Contributions with a photon-in-electron PDF (upper panel) and an electron-to-photon fragmentation function (lower panel).}
\end{figure*}

An interesting issue concerns the treatment of the electron. Clearly, at the energies we consider, the electron mass $m_e$ is expected to be negligible.
However, for vanishing $m_e$ one encounters additional singularities in the $2\to 3$ diagrams for $e q\to \gamma e q$, arising
when the initial or final electron is accompanied by collinear emission. One option is to keep a finite electron mass
throughout the full calculation, neglecting it wherever possible. The collinear singularities mentioned above would then
be replaced by finite (but, potentially large) logarithms of the form $\log(m_e/Q)$, with $Q$ a hard scale in the problem. Using this approach, the phase-space integrations with non-zero electron mass become slightly more involved. Alternatively -- and this is the approach we will show here -- one may introduce ``Weizs\"acker-Williams (WW)'' type distributions that correspond to a ``photon-in-electron'' PDF or an ``electron-to-photon'' fragmentation function, respectively, as shown in Fig.~\ref{fig:PDFWW}. The calculation may then be performed with $m_e=0$, regularizing collinear divergences associated with the electron by dimensional regularization and absorbing them into the electron distributions. As explained in Ref.~\cite{Hinderer:2015hra}, both approaches, i.e., with or without a finite electron mass, will yield equivalent results. 
Indeed, in addition to the WW approach, we have carried out the non-zero electron mass calculation and found complete agreement between the two methods. 
We also refer the reader to Refs.~\cite{Liu:2020rvc,Liu:2021jfp} for a discussion of radiative corrections in lepton scattering using
similar concepts. 

The second class of NLO contributions involves the hard-scattering functions for the processes $e q\to e qg$ and
$eg\to e q\bar{q}$, followed by fragmentation of a final-state parton to the photon. Using again that the photon fragmentation
functions are of order $\alpha_{\mathrm{em}}/\alpha_s$, the net contribution by these processes is of order
$\alpha_{\mathrm{em}}^2\alpha_s\times \alpha_{\mathrm{em}}/\alpha_s=\alpha_{\mathrm{em}}^3$, making them indeed NLO.
The corresponding calculations for $e q\to e qg$ and $eg\to e q\bar{q}$ where already carried out in Refs.~\cite{Hinderer:2015hra,Hinderer:2017ntk},
where inclusive-hadron production in $ep$ scattering was considered. The collinear singularities encountered in the
phase space integration for these reactions give rise to the homogeneous part of the evolution of the photon
fragmentation functions. Technically, this contribution can be obtained simply by replacing the parton-to-hadron fragmentation functions in the 
calculation of Refs.~\cite{Hinderer:2015hra,Hinderer:2017ntk} with the corresponding parton-to-photon ones. In the following we will
therefore not further discuss this contribution which will, however, be included in our numerical results below. 

We close this discussion with a note of caution. As we have described, we count the photon-in-proton PDF and the quark-to-photon fragmentation functions 
as LO, that is, ${\cal O}(\alpha_{\mathrm{em}}/\alpha_s)$. In contrast to this we regard the corresponding photon-in-{\it electron}
PDF and {\it electron}-to-photon FF as
${\cal O}(\alpha_{\mathrm{em}})$, despite the fact that they enter with an additional large logarithm $\log(m_e/Q)$. Although this
way of counting is arguably sensible from the point of view of QCD, it is clear that one should expect the contributions coming
with the leptonic
PDF and FF to be numerically as important as the ones entering with the photon-in-proton PDF and the quark-to-photon fragmentation functions.
Furthermore, the evolution of the latter resums large logarithms in the hard scale, while the logarithm $\log(m_e/Q)$ in the leptonic functions is only kept at fixed order. It would be possible to resum these logarithms as well to all orders
in $\alpha_{\mathrm{em}}$, but this is beyond the scope of the present work.

\subsection{\label{sub:ME} Parton Distribution Functions (PDFs) and Fragmentation Functions (FFs)}

Before proceeding to the actual calculation of the cross section to order $\mathcal{O}(\fsc^3)$, we 
collect all of the relevant hadronic/leptonic matrix elements and their parameterizations in terms of PDFs and FFs 
that we need for a factorized description of the cross section. 

The most common matrix elements that we encounter are those for the distribution functions for unpolarized or longitudinally polarized quarks in protons, 
$f_1^{q/p}(x,\mu)$ and $g_1^{q/p}(x,\mu)$, respectively. It is well-known that the two PDFs parameterize the following bilocal matrix element,
\begin{eqnarray}
&&\int_{-\infty}^\infty \tfrac{\mathrm{d}\lambda}{2\pi}\,\mathrm{e}^{i\lambda x}\langle p(P,S)|\,\bar{q}_j(0)\,\mathcal{W}[0;\lambda n]\,q_i(\lambda n)\,|p(P,S)\rangle\nonumber\\
&=&\frac12 \left(\slash{P}\,f_1^{q/p}(x,\mu)-S_L\,\slash{P}\gamma_5\,g_1^{q/p}(x,\mu)\right)_{ij}+...\,.\label{eq:DefPDF}
\end{eqnarray}
Note that this matrix element (\ref{eq:DefPDF}) incorporates a light-cone vector $n^\mu$ that satisfies $n^2=0$ and $P\cdot n=1$, but is otherwise arbitrary. In particular the Wilson line $\mathcal{W}[0;\lambda n]$ that renders the bilocal quark operator color-gauge invariant is along a straight line in the light-cone direction $n^\mu$. We explicitly keep chiral-even leading twist structures in the second line of (\ref{eq:DefPDF}), but neglect chiral-odd and/or subleading twist structures (indicated by the dots). We note that the apparent dependence on the explicit choice of the light-cone vector $n^\mu$ cancels out for the leading twist observables that we are interested in in this paper\footnote{The situation is more complicated for subleading-twist observables \cite{Kanazawa:2015ajw}.}. 
We also see that the first term in (\ref{eq:DefPDF}) is independent of the proton spin vector $S^\mu$ (with $S^2=-1$ and $P\cdot S=0$) while the second is proportional to the longitudinal projection of $S^\mu$ on $n^\mu$, i.e., $S_L=M\,(n\cdot S)$. In the approximation of a massless proton, $S_L$ may be considered as the proton's helicity. Lastly, we note that the matrix element (\ref{eq:DefPDF}) is to be considered a subtracted matrix element in which ultra-violet divergences are $\overline{\mathrm{MS}}$-subtracted. Then, the PDFs are to be read as $f_1^{q/p,\overline{\mathrm{MS}}}(x,\mu)$ 
and $g_1^{q/p,\overline{\mathrm{MS}}}(x,\mu)$, functions of the renormalization scale $\mu$. 

As discussed in the previous subsection, the process $e+p\to \gamma+X$ 
is also sensitive to the photon-in-proton 
distribution function $f_1^{\gamma/p}$ 
(and its polarized companion $g_1^{\gamma/p}$
), which enters the factorized description of the cross section already at LO. It is in fact the appearance of these lesser-known distribution functions that makes the process especially interesting.
The definition of the photon-in-proton distribution functions shares some similarities with that for the proton's gluon distribution function $f_1^{g/p}(x,\mu)$. To be specific, the unpolarized ($f_1^{\gamma/p}$) and polarized ($g_1^{\gamma/p}$) photon PDFs parameterize the following matrix element,
\begin{eqnarray}
&&\int_{-\infty}^\infty \tfrac{\mathrm{d}\lambda}{2\pi}\,\mathrm{e}^{i\lambda x}\langle p(P,S)|F^{n\nu}(0)\,F^{n\mu}(\lambda\,n)|p(P,S)\rangle\nonumber\\
&=& \frac{x}{2}\left[-\frac{g_{\perp}^{\mu\nu}}{1-\varepsilon}\,f_1^{\gamma/p}(x,\mu)+S_L\,i\epsilon^{Pn\nu\mu}\,g_1^{\gamma/p}(x,\mu)\right]\,.\label{eq:DefgammaPDF}
\end{eqnarray}

This matrix element (\ref{eq:DefgammaPDF}) incorporates the photonic gauge-invariant field-strength tensor $F^{\mu\nu}=\partial^\mu A^\nu-\partial^\nu A^\mu$, with short-hand notations of the form $F^{n\nu}=n_\mu F^{\mu\nu}$ etc. The photon field itself is represented by $A^\mu(x)$. The transverse projector $g_\perp^{\mu\nu}$ is given by $g_\perp^{\mu\nu}=g^{\mu\nu}-P^\mu n^\nu-P^\nu n^\mu$, and the antisymmetric tensor $\epsilon^{\mu\nu\rho\sigma}$ follows the convention $\epsilon^{0123}=+1$. We have anticipated the later use of dimensional regularization
with $d=4-2\varepsilon$ space-time dimensions. We note again that that photon PDFs in Eq.~(\ref{eq:DefgammaPDF}) can be related to the usual quark and gluon PDFs via the use of DIS structure functions~\cite{Manohar:2016nzj,Manohar:2017eqh,deFlorian:2024hsu}.

Furthermore, at order $\mathcal{O}(\fsc^3/\alpha_s)$, the photon can also be produced within a fragmentation process of a quark in the final state. This process is described by a quark-to-photon fragmentation matrix element,
\begin{eqnarray}
&&\frac{1}{N_c}\sumint \int_{-\infty}^\infty \tfrac{\mathrm{d}\lambda}{2\pi}\,\mathrm{e}^{i\frac{\lambda}{z}}\langle 0 |\,\mathcal{W}[\infty n;\lambda n]q_i(\lambda n)\,|\gamma(P_\gamma);X\rangle\times\nonumber\\
&&\hspace{3cm}\langle \gamma(P_\gamma);X|\,\bar{q}_j(0)\,\mathcal{W}[0;\infty n]\,|0\rangle\nonumber\\
&=& z^{-1+2\varepsilon} (\slash{P}_\gamma)_{ij}\,D_1^{\gamma/q}(z,\mu)\,.\label{eq:DefFF}
\end{eqnarray}
The quark-to-photon fragmentation functions $D_1^{\gamma/q}$, like the photonic PDF counterparts, are only poorly known, despite their
relevance for direct-photon production at hadron colliders. Hence, measurements of  $e+p\to \gamma+X$ cross sections could potentially further constrain these fragmentation functions.

Like the quark PDFs (\ref{eq:DefPDF}) and any collinear matrix element, also the photon PDFs/FFs (\ref{eq:DefgammaPDF}) and (\ref{eq:DefFF}) contain ultraviolet divergences that need to be regularized and subtracted, thereby introducing scale dependence. This is a standard procedure, and a convenient way is to consider perturbative corrections to the collinear functions and to introduce ``renormalized" $\overline{\mathrm{MS}}$-PDFs/FFs. To the order we consider, we have
\begin{eqnarray}
f_1^{\gamma/p}(x,\mu) & = & f_1^{\gamma/p,\overline{\mathrm{MS}}}(x,\mu)+ \frac{\fsc}{2\pi}\frac{S_\varepsilon}{\varepsilon}\times\label{eq:Renfgamma}\nonumber\\
&&\hspace{-2cm}\int_x^1\tfrac{\mathrm{d}w}{w}\,P_{\gamma q}(w)\sum_q e_q^2 \,f_1^{(q+\bar{q})/p,\overline{\mathrm{MS}}}(x/w,\mu)
,\\
g_1^{\gamma/p}(x,\mu) & = & g_1^{\gamma/p,\overline{\mathrm{MS}}}(x,\mu)+ \frac{\fsc}{2\pi}\frac{S_\varepsilon}{\varepsilon}\times \label{eq:renggamma}\nonumber\\
&&\hspace{-2cm}\int_x^1\tfrac{\mathrm{d}w}{w}\Delta P_{\gamma q}(w)\sum_q e_q^2 \,g_1^{(q+\bar{q})/p,\overline{\mathrm{MS}}}(x/w,\mu)
,\\
D_1^{\gamma / q}(z,\mu) & = & D_1^{\gamma / q,\overline{\mathrm{MS}}}(z,\mu)+ \frac{\fsc}{2\pi}\frac{S_\varepsilon}{\varepsilon}e_q^2\,P_{\gamma q}(z)\label{eq:renFF}
\,.
\end{eqnarray}
The kernels in these equations are the LO QED 
splitting functions $P_{\gamma q}(w)=\frac{1+(1-w)^2}{w}$ and $\Delta P_{\gamma q}(w)=2-w$. The customary factor $S_\varepsilon=(4\pi)^\varepsilon/\Gamma(1-\varepsilon)$ facilitates the correct $\overline{\mathrm{MS}}$ subtraction at this order. In (\ref{eq:Renfgamma}),(\ref{eq:renggamma}) the sum is over quark flavors, $e_q$ are the fractional quark charges, and $f^{(q+\bar{q})/p}$ 
indicates a sum of quark and anti-quark distributions. Finally, the last term in (\ref{eq:renFF}) indicates the perturbative radiation of a collinear photon that is part of the quark-to-photon fragmentation.

As described earlier, it turns out to be convenient to similarly introduce photon PDFs for an {\it electron} 
rather than a proton 
and, likewise, electron-to-photon 
FFs. To be specific, we may define photon-in-electron 
distributions $f_1^{\gamma/e}(x,\mu)$
, $g_1^{\gamma/e}(x,\mu)$ 
in the same way as the photon-in-proton 
distributions by replacing the proton 
state $p(P,S)$ 
by an electron 
state $e(l,s)$
. The main difference is that the $f_1^{\gamma/e}(x,\mu)$
, $g_1^{\gamma/e}(x,\mu)$ (sometimes referred to as "Weisz\"acker-Williams" (WW) distributions \cite{vonWeizsacker:1934nji,Williams:1934ad}) can be calculated perturbatively for a non-zero electron mass $m_e$, and that they do not couple to the quark distributions. The corresponding calculations to order $\mathcal{O}(\fsc)$ have been performed in Refs.~\cite{Hinderer:2015hra,Hinderer:2017ntk}, and we simply quote the results:
\begin{eqnarray}
f_1^{\gamma/e}(x,\mu) &=& f_1^{\gamma/e,\overline{\mathrm{MS}}}(x,\mu)+ \frac{\fsc}{2\pi}\frac{S_\varepsilon}{\varepsilon}\,P_{\gamma q}(x)+\mathcal{O}(\fsc^2)\,,\label{eq:renWWf1}\\
g_1^{\gamma/e}(x,\mu) &=& g_1^{\gamma/e,\overline{\mathrm{MS}}}(x,\mu)+ \frac{\fsc}{2\pi}\frac{S_\varepsilon}{\varepsilon}\,\Delta P_{\gamma q}(x)+\mathcal{O}(\fsc^2)\,,\label{eq:renWWg1}
\end{eqnarray}
with
\begin{eqnarray}
f_1^{\gamma/e,\overline{\mathrm{MS}}}(x,\mu) & = & \frac{\fsc}{2\pi}\,P_{\gamma q}(x)\,\left[\ln\left(\frac{\mu^2}{x^2\,m_e^2}\right)-1\right]\,,\label{eq:f1WW}\\
g_1^{\gamma/e,\overline{\mathrm{MS}}}(x,\mu) & = & \frac{\fsc}{2\pi}\,\Delta P_{\gamma q}(x)\,\ln\left(\frac{\mu^2}{x^2\,m_e^2}\right)\,.\label{eq:g1WW}
\end{eqnarray}
We note that detailed studies on the evolution of the photon-in-lepton distributions have been presented recently in Refs.~\cite{Frixione:2012wtz,Bertone:2019hks,Frixione:2019lga}.\\

In similar fashion, we also introduce an electron-to-photon 
fragmentation function $D_1^{\gamma/e}(z)$ 
by replacing the quark fields in (\ref{eq:DefFF}) by electron fields. As for the electron 
PDFs, we can calculate all involved quantities in QED perturbation theory, and obtain
\begin{eqnarray}
D_1^{\gamma / e}(z,\mu)  =  D_1^{\gamma / e,\overline{\mathrm{MS}}}(z,\mu)+ \frac{\fsc}{2\pi}\frac{S_\varepsilon}{\varepsilon}\,\,P_{\gamma q}(z)+\mathcal{O}(\fsc^2)\,,\label{eq:renFFWW}
\end{eqnarray}
with
\begin{eqnarray}
D_1^{\gamma/e,\overline{\mathrm{MS}}}(z,\mu) & = & \frac{\fsc}{2\pi}\,P_{\gamma q}(z)\,\left[\ln\left(\frac{\mu^2}{z^2\,m_e^2}\right)-1\right]\,.\label{eq:D1WW}
\end{eqnarray}
We note that leptonic PDFs have also been introduced in more general terms 
in the context of factorized expressions for QED radiative corrections to the DIS and semi-inclusive DIS cross sections~\cite{Liu:2021jfp}.

\subsection{\label{sub:LO}LO partonic cross sections}

We introduce the Mandelstam variables $s=(P+l)^2>0$, $t=(P-P_\gamma)^2<0$ and $u=(l-P_\gamma)^2<0$, which satisfy $s+t+u \ge 0$. 
We further define 
\begin{eqnarray}
x_0=\frac{-u}{s+t}\,,\quad\quad v=\frac{s+t}{s}<1\,. 
\end{eqnarray}

We start by analyzing the contribution to the cross section associated with the photon PDF. As indicated in the upper panel of Fig.~\ref{fig:LO},
we factorize this contribution into $f_1^{\gamma/p}$ and the hard-scattering function for QED Compton scattering $\gamma e\to\gamma e$, which may computed from the diagrams shown in the figure. Although the result is finite at this order, we compute it in dimensional
regularization with $d=4-2\varepsilon$ space-time dimensions. The reason is that we need the $d$-dimensional cross section later in the factorization
of the $2\to 3$ scattering contributions. A straightforward calculation gives 
\begin{eqnarray}
E_\gamma \frac{\mathrm{d}^{d-1}\sigma^{\gamma \mathrm{PDF}}}{\mathrm{d}^{d-1}\bm{P}_\gamma} &=& \frac{\fsc^2}{s(-u)}\,\hat{\sigma}^{\gamma\mathrm{PDF}}(v,\varepsilon)\,f_1^{\gamma/p}(x_0,\mu)\,,\label{eq:CSgammaPDF}
\end{eqnarray}
with
\begin{eqnarray}
\hat{\sigma}^{\gamma\mathrm{PDF}}(v,\varepsilon) & = & 2\frac{1+v^2}{v}-2\varepsilon\frac{(1-v)^2}{v}\,.\label{eq:partCDgammaPDF}
\end{eqnarray}
Here $\bm{P}_\gamma$ denotes the three-momentum of the produced photon. 
If (\ref{eq:CSgammaPDF}) were the only contribution to $e N\to \gamma X$, the process would be ideally suited to constrain the photon-in-proton 
PDF $f_1^{\gamma/p}$.

However, as already discussed in subsection~\ref{sub:LOandNLO}, there are also LO contributions associated with quark-to-photon
fragmentation. The corresponding factorization is sketched in the lower part of Fig.~\ref{fig:LO}. We find
\begin{eqnarray}
E_\gamma \frac{\mathrm{d}^{d-1}\sigma^{\gamma \mathrm{FF}}}{\mathrm{d}^{d-1}\bm{P}_\gamma} &=&\frac{\fsc^2}{s(-u)}\int_{x_0}^1\tfrac{\mathrm{d}w}{w}\,\hat{\sigma}^{\gamma\mathrm{FF}}(v,w,\varepsilon)\times\label{eq:CSgammaFF}\nonumber\\
&&\sum_q e_q^2\,f_1^{(q+\bar{q})/p}\left(\tfrac{x_0}{w},\mu\right)\,D_1^{\gamma/q}(1-v+vw,\mu)\,,
\end{eqnarray}
where
\begin{eqnarray}
\hat{\sigma}^{\gamma\mathrm{FF}}(v,w,\varepsilon) &=& \frac{2vw(1-v+vw)^{2\varepsilon}}{(1-v)^2(1-v+vw)}\times\label{eq:partCSgammaFF}\nonumber\\
&&\hspace{-1cm}\left[1+v^2-2v(1-w)(1+vw)-\varepsilon (1-v)^2\right].
\end{eqnarray}
As indicated in~(\ref{eq:CSgammaFF}), quarks and antiquarks have identical fragmentation functions to photons.

\subsection{\label{sub:NLO}Contributions at $\mathcal{O}(\fsc^3)$}

In this subsection we discuss the contributions sketched in Figs.~\ref{fig:PM} and~\ref{fig:PDFWW}. We first address the $2\to 3$ 
diagrams shown in Fig.~\ref{fig:PM}, for which the observed photon can be radiated either from the electron or from the quark line. 
Radiation purely off the electron is known as the \textit{Bethe-Heitler} (BH) contribution~\cite{Brodsky:1972yx,Albaltan:2019cyc},
while the diagrams with radiation off quarks may be regarded as a quark-photon \textit{Compton} (C) contribution. 
Finally, there is also an \textit{interference} (I) contribution of the two types of radiation.  

The three types of contributions,  BH,C,I, can be distinguished by the weightings with different quark charges. 
Furthermore, the Bethe-Heitler and Compton contributions come with the sums of quark and anti-quark distributions, while the interference contribution 
is generated by the valence quark distributions. To be specific, we encounter the following generic combinations of quark PDFs \cite{Albaltan:2019cyc}:
\begin{eqnarray}
f^{{\textrm{BH}}}(x,\mu)&=&\sum_q e_q^2\big(f^{q/p}(x,\mu)+f^{\bar{q}/p}(x,\mu)\big)\,,\label{eq:PDFBH}\\
f^{{\textrm{C}}}(x,\mu) & = & \sum_q e_q^4 \big(f^{q/p}(x,\mu)+f^{\bar{q}/p}(x,\mu)\big)\,,\label{eq:PDFC}\\
f^{{\textrm{I}}}(x,\mu) & = & \sum_q e_q^3 \big(f^{q/p}(x,\mu)- f^{\bar{q}/p}(x,\mu)\big)\,.\label{eq:PDFI}
\end{eqnarray}
In order to obtain the hard-scattering cross section for $qe \to qe\gamma$ we need to integrate its squared amplitude over the 
phase space of the momenta of the undetected electron and quark ($r_1$ and $r_2$ in Fig.~\ref{fig:PM}), which is done in 
$d=4-2\varepsilon$ dimensions. As mentioned earlier, we use a vanishing electron mass $m_e=0$ to simplify the calculation.
Although the phase space integrations become quite standard then, we found the recent paper \cite{Lyubovitskij:2021ges}
particularly helpful in this context. After performing the phase-space integrations we encounter various types of $1/\varepsilon$-poles associated
with the Bethe-Heitler and Compton contributions, arising when the radiated photon becomes either collinear to the quark or the electron momentum.  
In contrast to the Bethe-Heitler and Compton contributions, the interference contributions are finite. 

The collinear divergences in the Bethe-Heitler and Compton contributions can be systematically absorbed into 
the hadronic and leptonic PDFs and FFs. For initial-state radiation off the quark, the collinear pole is canceled
by that in the perturbative photon PDF in~(\ref{eq:Renfgamma}), inserted into the LO cross section in~(\ref{eq:CSgammaPDF}). 
Likewise, the final-state collinear singularity cancels against the quark-to-photon fragmentation contribution~(\ref{eq:renFF}),
which appears via the LO term~(\ref{eq:partCSgammaFF}). In both cases, we perform the subtraction in the $\overline{\mathrm{MS}}$
scheme. 

For collinear radiation off the electron line, the situation is slightly less standard. As discussed in subsection~\ref{sub:LOandNLO},
our approach is to set $m_e=0$, in which case collinear initial-state and final-state divergences will occur. We subsequently add perturbative 
``Weizs\"{a}cker-Williams'' contributions associated with a photon-in-electron PDF and an electron-to-photon FF, which cancel the 
divergences and ``reinstate'' the explicit logarithms $\log(m_e/Q)$
that would arise in a calculation that keeps a finite electron mass. Expressions for the relevant PDF and FF were given in
Eqs.~(\ref{eq:renWWf1}) and~(\ref{eq:renFFWW}), respectively. They need to be convoluted with the appropriate Born
processes which are $\gamma q\to \gamma q$ in the PDF case and $e q\to e q$ in the FF one. We obtain:
\begin{eqnarray}
E_\gamma \frac{\mathrm{d}^3\sigma^{\gamma \mathrm{PDF}}_{\mathrm{WW}}}{\mathrm{d}^{d-1}\bm{P}_\gamma} &=& \frac{\fsc^2}{s(-u)}\int_{x_0}^1\tfrac{\mathrm{d}w}{w} \,\hat{\sigma}^{\gamma\mathrm{PDF}}_{\mathrm{WW}}(v,w,\varepsilon)\times\,\label{eq:CSgammaPDFWW}\nonumber\\
& & f_1^{\gamma/e}\left(\tfrac{1-v}{1-vw},\mu\right)\,\sum_q e_q^4\,f_1^{(q+\bar{q})/p}\left(\tfrac{x_0}{w},\mu\right)\,,
\end{eqnarray}
where
\begin{eqnarray}
\hat{\sigma}^{\gamma\mathrm{PDF}}_{\mathrm{WW}}(v,w,\varepsilon) &=&\frac{2vw(2-vw(2-vw(1-\varepsilon)))}{(1-v)(1-vw)}\,,\label{eq:partCSgammaPDFWW}
\end{eqnarray}
and 
\begin{eqnarray}
E_\gamma \frac{\mathrm{d}^3\sigma^{\gamma \mathrm{FF}}_{\mathrm{WW}}}{\mathrm{d}^{d-1}\bm{P}_\gamma} &=&\frac{\fsc^2}{s(-u)}\int_{x_0}^1\tfrac{\mathrm{d}w}{w}\,\hat{\sigma}^{\gamma\mathrm{FF}}_{\mathrm{WW}}(v,w,\varepsilon)\times\label{eq:CSgammaFFWW}\nonumber\\
&&D_1^{\gamma/e}(1-v+vw,\mu)\sum_q e_q^2\,f_1^{(q+\bar{q})/p}\left(\tfrac{x_0}{w},\mu\right)\,,
\end{eqnarray}
with 
\begin{eqnarray}
\hat{\sigma}^{\gamma\mathrm{FF}}_{\mathrm{WW}}(v,w,\varepsilon) &=& \frac{2(1-v+vw)^{2\varepsilon}}{vw(1-v+vw)}\times\label{eq:partCSgammaFFWW}
\nonumber\\
&&\hspace{-1cm}\left[2(1-v)(1-v+vw)+(1-\varepsilon)\,v^2w^2\right].
\end{eqnarray}
As mentioned before, resummation of the logarithm $\log(m_e/Q)$ may eventually be necessary.

\subsection{\label{sub:ResCS}Final result for the unpolarized cross section}

We separate the invariant cross section for $e(l)+p(P)\to \gamma(P_\gamma)+X$ into five pieces: the 
contributions associated with (1) the photon-in-proton PDF and (2) the quark-to-photon FF (which both start at $\mathcal{O}(\fsc^2/\alpha_s)$),
and the ones by the (3) Bethe-Heitler, (4) Compton, (5) interference channels, which all arise only at $\mathcal{O}(\fsc^3)$ and are hence NLO:
\begin{eqnarray}
E_\gamma \frac{\mathrm{d}^3\sigma^{e p\to \gamma X}}{\mathrm{d}^{3}\bm{P}_\gamma}&\equiv& 
 \sigma^{\gamma\mathrm{PDF}}+\sigma^{\gamma\mathrm{FF}}+ \sigma^{\mathrm{BH}}+ \sigma^{\mathrm{C}}+ \sigma^{\mathrm{I}}
\,.\label{eq:CSDecomp}
\end{eqnarray}
We emphasize again that the first contribution in (\ref{eq:CSDecomp}) is associated with 
the photon-in-proton PDF $f_1^{\gamma/p}$ while the second term is generated by the quark-to-photon fragmentation functions $D_1^{\gamma/q}$. 
We find
\begin{eqnarray}
\hspace{-0.5cm}\sigma^{\gamma\mathrm{PDF}} &=& \frac{2\fsc^2}{s(-u)}\,\hat{\sigma}^{\gamma\mathrm{PDF}}(v)\,f_1^{\gamma/p,\overline{\mathrm{MS}}}(x_0,\mu)\,,\label{eq:resCSgammaPDF}\\
\sigma^{\gamma\mathrm{FF}} & = & \frac{2\fsc^2}{s(-u)}\,\int_{x_0}^1\tfrac{\mathrm{d}w}{w}\,\hat{\sigma}^{\gamma\mathrm{FF}}(v,w)\times\label{eq:resCSgammaFF}\nonumber\\
&& \hspace{-2cm}\sum_q e_q^2\,f_1^{(q+\bar{q})/p,\overline{\mathrm{MS}}}\left(\tfrac{x_0}{w},\mu\right)\,
D_1^{\gamma/q,\overline{\mathrm{MS}}}(1-v+vw,\mu)\,,\\
\sigma^{\mathrm{BH}}& = & \frac{\fsc^3}{\pi \,s(-u)}\int_{x_0}^1\tfrac{\mathrm{d}w}{w}\,\hat{\sigma}^{\mathrm{BH}}\left(v,w,\tfrac{-u}{m_e^2},\tfrac{-u}{\mu^2}\right)\times\nonumber\\
&&\hspace{2.5cm} f_1^{\mathrm{BH},\overline{\mathrm{MS}}}\left(\tfrac{x_0}{w},\mu\right)\,,\label{eq:resCSBH}\\
 \sigma^{\mathrm{C}}& = &\frac{\fsc^3}{\pi \,s(-u)}\int_{x_0}^1\tfrac{\mathrm{d}w}{w}\,\hat{\sigma}^{\mathrm{C}}\left(v,w,\tfrac{-u}{m_e^2},\tfrac{-u}{\mu^2}\right)\times\nonumber\\
&&\hspace{2.5cm} f_1^{\mathrm{C},\overline{\mathrm{MS}}}\left(\tfrac{x_0}{w},\mu\right)\,,\label{eq:resCSC}\\
\sigma^{\mathrm{I}}& = &\frac{\fsc^3}{\pi \,s(-u)}\int_{x_0}^1\tfrac{\mathrm{d}w}{w}\,\hat{\sigma}^{\mathrm{I}}(v,w)\, f_1^{\mathrm{I},\overline{\mathrm{MS}}}\left(\tfrac{x_0}{w},\mu\right)\,.\label{eq:resCSI}
\end{eqnarray}
The various partonic hard-scattering functions $\hat\sigma$ appearing in these equations are collected in Appendix \ref{app:UCS}. 

\subsection{\label{sub:Isolation} Photon isolation}

In collider experiments, in order to cope with the large photon background arising from neutral-pion decay, 
one usually introduces an \textit{isolation} of the photon, for which one imposes a limit on the hadronic activity in the vicinity of the photon. 
Typically one defines a cone in azimuthal-pseudorapidity space (as defined in the electron-proton c.m. frame) around the photon candidate, i.e.
\begin{equation}
    \mathrm{cone}_\gamma(R)=\left\{(\eta,\phi)|\sqrt{(\eta-\eta_\gamma)^2+(\phi-\phi_\gamma)^2}\le R \right\}\,.\label{eq:DefCone}
\end{equation}
Inside this cone one then requires that the energy of any hadrons accompanying the photon is smaller than a fixed fraction
of the photon energy, i.e.
\begin{equation}
    E_{\textrm{had}}\le \xi E_\gamma\,.\label{eq:IsoCrit}
\end{equation}
where $\xi \sim 0.1$ is a typical value. Anticipating that isolation would also be imposed in photon measurements
at the EIC, we will extend our calculation to the isolated case. Actually, a benefit of isolation is that it also
reduces the contribution of photons that were generated in fragmentation of quarks or gluons. The reason is
that for the fragmentation component photons will indeed always be accompanied by hadronic energy,
so that the isolation condition~(\ref{eq:IsoCrit}) puts restrictions on the size of the contributions, although it 
obviously does not eliminate it completely. To be more specific, one can see that 
criterion (\ref{eq:IsoCrit}) translates to a condition on the light-cone fraction $z=1-v+vw$ that appears in the fragmentation 
contribution (\ref{eq:resCSgammaFF})~\cite{Berger:1990es}:
\begin{eqnarray}
   E_{\textrm{had}} \le \xi E_\gamma  & \Leftrightarrow & z\ge \frac{1}{1+\xi}\nonumber\\
   &\Leftrightarrow & w \ge \mathrm{max}\left(x_0,1-\frac{\xi}{v(1+\xi)}\right)\equiv w_0\,.\label{eq:Defw0}
\end{eqnarray}
Thus the integration domain over the variable $w$ in (\ref{eq:resCSgammaFF}) is reduced, and 
the smaller the fraction $\xi$ the stronger is the suppression of the parton-to-photon fragmentation.

Beyond LO, partons can also be radiated into the photon isolation cone, rather than just enter it
via fragmentation. This effect must be taken into account in the ${\cal O}(\alpha_{\mathrm{em}}^3)$
contribution. In order to do this, we adopt the \textit{small cone approximation} (SCA), see Ref.~\cite{Gordon:1994ut}. 
The basic idea is to use the results obtained for the fully inclusive (i.e., non-isolated) prompt photon cross section, Eqs.
(\ref{eq:resCSBH})--(\ref{eq:resCSI}), and subtract the NLO contribution where partons have entered the isolation
cone with energy in excess of what is allowed by the condition~(\ref{eq:IsoCrit}). For small cone radii $R$ the subtraction contribution defined
in this way may be computed analytically. More specifically, for the case of interest here it can be shown to have the form 
\begin{equation}
\sigma_{\mathrm{sub}}=\ln(R)\,A+B+\mathcal{O}(R^2)\,.\label{eq:IsoSub}
\end{equation}
The two leading terms associated with $A$ and $B$ can only arise from configurations where the photon is radiated
by an outgoing quark or antiquark. For nearly collinear emission, the quark propagator before emission goes nearly on-shell,
so that phase space integration leads to the structure of logarithm plus constant shown in~(\ref{eq:IsoSub}). 
The coefficients $A$ and $B$ contain a Born cross section which in this case is the one for the Compton 
process $e q\to e q$. The explicit calculation of the full subtraction term in the SCA is straightforward, and for details we refer the reader to 
Ref.~\cite{Gordon:1994ut}. In the end we obtain 
\begin{eqnarray}
\sigma^{\mathrm{C}}_{\mathrm{sub}}& = & \theta(w_0-x_0)\frac{\fsc^3}{\pi \,s(-u)}\int_{x_0}^{w_0}\tfrac{\mathrm{d}w}{w}\,\times\nonumber\\
&&\hspace{-1.5cm} \hat{\sigma}^{\mathrm{C}}_{\mathrm{sub}}\left(v,w,x_0,R,\tfrac{-u}{\mu^2}\right)\,f_1^{\mathrm{C},\overline{\mathrm{MS}}}\left(\tfrac{x_0}{w},\mu\right)\,,\label{eq:resCSsub}
\end{eqnarray}
where
\begin{eqnarray}
    \hat{\sigma}^{\mathrm{C}}_{\mathrm{sub}}\left(v,w,x_0,R,\tfrac{-u}{\mu^2}\right) & = & -\frac{vw}{(1-v)^2(1-v+vw)^2}\times\nonumber\\
    && \hspace{-1.4cm}(1+v^2-2v(1-w)(1+vw))\,\times\nonumber\\
    && \hspace{-1.5cm}\left[(1-v+vw)^2+(1+v^2(1-w)^2)\times\right.\nonumber\\
    && \hspace{-3.3cm}\left.\ln\left(-\frac{R^2\,u\,v(1-v)(1-w)(1-v+vw)^2x_0}{\mu^2\,w(1-v+vx_0)^2}\right)\right]\,.\label{eq:partresCSsub}
\end{eqnarray}
The term in~(\ref{eq:resCSsub}) is to be subtracted from the Compton contribution in~(\ref{eq:CSDecomp}). 
Empirically, the SCA is known to work well for values of $R$ up to $R\sim 0.7$~\cite{Gordon:1994ut}. Note that for larger values of $R$ terms of order $\mathcal{O}(R^2)$ in Eq.~(\ref{eq:IsoSub}) are expected to become relevant. Such terms are beyond the scope of the SCA. On the other hand, if very small cone radii $R$ are considered a resummation of the $\log(R)$-term in Eq.~(\ref{eq:IsoSub}) may be required \cite{Kang:2016mcy}.

As we discussed earlier, there is a second class of NLO contributions to $e(l)+p(P)\to \gamma(P_\gamma)+X$ which is
generated by the hard-scattering processes $e q\to e qg$ and $eg\to e q\bar{q}$, followed by fragmentation of a final-state parton to the photon. 
In principle, we would need to introduce appropriate subtraction terms also for this contribution. 
However, in our numerical studies to be presented below we found that this contribution is strongly reduced already by the condition~(\ref{eq:Defw0}),
so that it becomes irrelevant. We therefore refrain from deriving the subtraction piece for this part. 

We finally note that the Bethe-Heitler and interference contributions only contribute to the terms of order $R^2$ in~(\ref{eq:IsoSub}). 
The reason for this is that we only impose isolation in terms of {\it hadronic} energy. In principle,
one could introduce also an ``electromagnetic isolation'', restricting the amount of energy of additional electromagnetic
particles around the photon. Such an isolation would affect especially the size of the Bethe-Heitler contribution,
where it would then also generate logarithms of the cone size. 
Whether or not such a constraint would be desirable (or even feasible experimentally) remains to be studied.

\section{\label{sec:LCS}The polarized cross section}

In this section we extend our calculation to photon production in collisions of longitudinally polarized electrons and protons. 
To be specific, we consider the following spin dependent cross section:
\begin{eqnarray}
    \Delta \sigma^{e p\to \gamma X}&\equiv  &\frac14 \left[
    E_\gamma \frac{\mathrm{d}^3\sigma^{e^+ p^+\to\gamma X}}{\mathrm{d}^3\bm{P}_\gamma}
    -E_\gamma \frac{\mathrm{d}^3\sigma^{e^- p^+\to\gamma X}}{\mathrm{d}^3\bm{P}_\gamma}\right.\nonumber\\
    &&\left.-E_\gamma \frac{\mathrm{d}^3\sigma^{e^+ p^-\to\gamma X}}{\mathrm{d}^3\bm{P}_\gamma}+
    E_\gamma \frac{\mathrm{d}^3\sigma^{e^- p^-\to\gamma X}}{\mathrm{d}^3\bm{P}_\gamma}\right]\,,
    \label{eq:LCS}
\end{eqnarray}
where the superscripts refer to the electron and proton helicities. Since $\--$ like the unpolarized cross section $\--$ also 
$\Delta \sigma$ is a leading-power observable, the discussions of subsections \ref{sub:ME} - \ref{sub:NLO} will hold for $\Delta \sigma$ as well, and the calculations can be performed in the same fashion. A slight complication compared to the unpolarized cross section is the treatment of the Dirac matrix $\gamma_5$ in dimensional regularization which arises from projecting on longitudinal polarization of the electron and the proton. In this work we choose to work within the \textit{'t Hooft-Veltman-Breitenloher-Maison} scheme for $\gamma_5$ \cite{tHooft:1972tcz,Breitenlohner:1977hr}.

As before, we split the spin dependent cross section into five pieces (cf. Eq.~(\ref{eq:CSDecomp})),
\begin{eqnarray}
    \Delta \sigma^{e p\to \gamma X} &=&  \Delta \sigma^{\gamma\mathrm{PDF}}+\Delta \sigma^{\gamma\mathrm{FF}}
    +\Delta \sigma^{\mathrm{BH}}+\Delta \sigma^{\mathrm{C}}+\Delta \sigma^{\mathrm{I}}.\label{eq:LCSDecomp}
\end{eqnarray}
The individual contributions are as follows:
\begin{eqnarray}
\hspace{-0.5cm}\Delta \sigma^{\gamma\mathrm{PDF}} &=& \frac{2\fsc^2}{s(-u)}\,\Delta\hat{\sigma}^{\gamma\mathrm{PDF}}(v)\,g_1^{\gamma/p,\overline{\mathrm{MS}}}(x_0,\mu)\,,\label{eq:resLCSgammaPDF}\\
\Delta \sigma^{\gamma\mathrm{FF}} & = & \frac{2\fsc^2}{s(-u)}\,\int_{x_0}^1\tfrac{\mathrm{d}w}{w}\,\Delta\hat{\sigma}^{\gamma\mathrm{FF}}(v,w)\times\label{eq:resLCSgammaFF}\nonumber\\
&& \hspace{-2cm}\sum_{q}e_q^2\,g_1^{(q+\bar{q})/p,\overline{\mathrm{MS}}}\left(\tfrac{x_0}{w},\mu\right)\,D_1^{\gamma/q,\overline{\mathrm{MS}}}(1-v+vw,\mu)\,,\\
\Delta \sigma^{\mathrm{BH}} & = & \frac{\fsc^3}{\pi \,s(-u)}\int_{x_0}^1\tfrac{\mathrm{d}w}{w}\,\Delta\hat{\sigma}^{\mathrm{BH}}\left(v,w,\tfrac{-u}{m_e^2},\tfrac{-u}{\mu^2}\right)\times\nonumber\\
&&\hspace{2.5cm} g_1^{\mathrm{BH},\overline{\mathrm{MS}}}\left(\tfrac{x_0}{w},\mu\right)\,,\label{eq:resLCSBH}\\
\Delta \sigma^{\mathrm{C}} & = &\frac{\fsc^3}{\pi \,s(-u)}\int_{x_0}^1\tfrac{\mathrm{d}w}{w}\,\Delta\hat{\sigma}^{\mathrm{C}}\left(v,w,\tfrac{-u}{m_e^2},\tfrac{-u}{\mu^2}\right)\times\nonumber\\
&&\hspace{2.5cm} g_1^{\mathrm{C},\overline{\mathrm{MS}}}\left(\tfrac{x_0}{w},\mu\right)\,,\label{eq:resLCSC}\\
\Delta \sigma^{\mathrm{I}} & = &\frac{\fsc^3}{\pi \,s(-u)}\int_{x_0}^1\tfrac{\mathrm{d}w}{w}\,\Delta\hat{\sigma}^{\mathrm{I}}(v,w)\, g_1^{\mathrm{I},\overline{\mathrm{MS}}}\left(\tfrac{x_0}{w},\mu\right)\,.\label{eq:resLCSI}
\end{eqnarray}
We observe that the helicity distributions $g_1^{q/p}$, $g_1^{\gamma/p}$, $g_1^{\gamma/e}$ (see Eqs.~(\ref{eq:DefPDF}), (\ref{eq:DefgammaPDF})) appear in these expressions. The explicit analytic forms of the polarized hard-scattering functions $\Delta \hat{\sigma}$ in Eqs.~(\ref{eq:resLCSgammaPDF})--(\ref{eq:resLCSI}) are presented in Appendix \ref{app:LCS}. 

In order to construct the spin-dependent cross section for isolated-photon production we repeat the steps described in the previous section and again compute 
the relevant subtraction term for the polarized Compton contribution (\ref{eq:resLCSC}):
\begin{eqnarray}
    \Delta \sigma^{\mathrm{C}}_{\mathrm{sub}}& = &\theta(w_0-x_0)\frac{\fsc^3}{\pi \,s(-u)}\int_{x_0}^{w_0}\tfrac{\mathrm{d}w}{w}\,\times\nonumber\\
&&\hspace{-1.5cm}\Delta\hat{\sigma}^{\mathrm{C}}_{\mathrm{sub}}\left(v,w,x_0,R,\tfrac{-u}{\mu^2}\right)\, g_1^{\mathrm{C},\overline{\mathrm{MS}}}\left(\tfrac{x_0}{w},\mu\right)\,,\label{eq:resLCSsub}
\end{eqnarray}
with 
\begin{eqnarray}
    \Delta\hat{\sigma}^{\mathrm{C}}_{\mathrm{sub}}\left(v,w,x_0,R,\tfrac{-u}{\mu^2}\right) & = & -\frac{vw(1-v+2vw)}{(1-v)(1-v+vw)^2}\times\nonumber\\
    && \hspace{-1.5cm}\left[(1-v+vw)^2+(1+v^2(1-w)^2)\times\right.\nonumber\\
    && \hspace{-3.3cm}\left.\ln\left(-\frac{R^2\,u\,v(1-v)(1-w)(1-v+vw)^2x_0}{\mu^2\,w(1-v+vx_0)^2}\right)\right]\,.\label{eq:partresLCSsub}
\end{eqnarray}

\section{\label{sec:Numerics}Numerical Predictions}

In this section we estimate the size of the various contributions to the unpolarized and polarized cross
sections at the EIC. Our main goal is to investigate in how far the future EIC experiments could help to probe the
proton's photon PDF. 

Throughout our calculations we use the NLO unpolarized parton distributions of Ref.~\cite{Martin:2009iq}, referred to as MSTW2008. For 
the helicity parton distributions we use the latest NLO set of Ref.~\cite{deFlorian:2014yva} (DSSV). For the parton-to-photon fragmentation functions we adopt 
the GRV NLO set \cite{Gluck:1992zx}. The photon-in-proton PDF 
is implemented using a recent NLO fit performed by the LuxQED collaboration \cite{Bertone:2017bme} and included in the NNPDF3.1 parameterizations. 

\begin{figure*}
\includegraphics[width=1\textwidth]{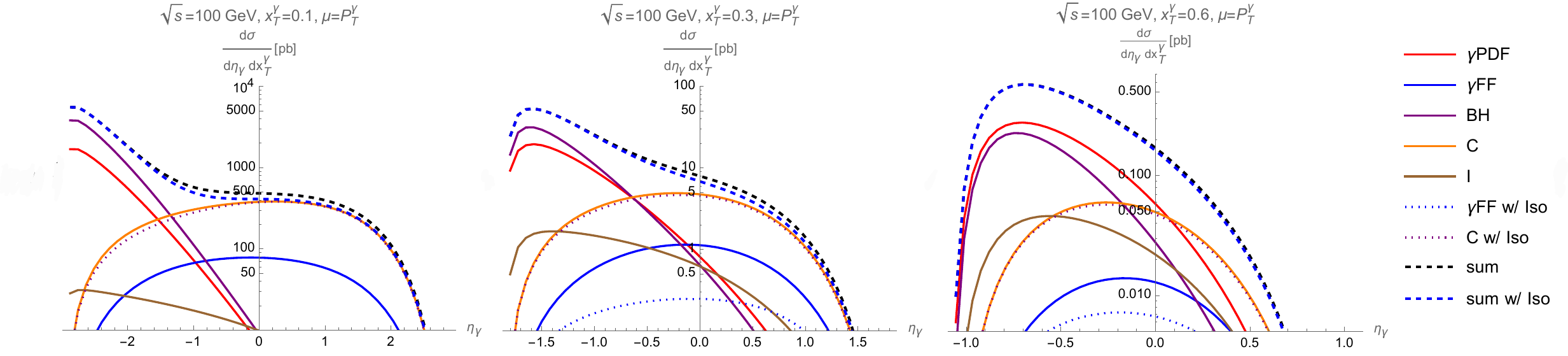}
\caption{\label{fig:CSEICeta} The unpolarized cross section at the EIC for c.m.-energy $\sqrt{s}=100\,\mathrm{GeV}$, plotted as function of the photon's pseudorapidity $\eta_\gamma$ for fixed transverse momenta. Left: $p_T^\gamma=5\,\mathrm{GeV}$; center: $p_T^\gamma=15\,\mathrm{GeV}$; right: 
$p_T^\gamma=30\,\mathrm{GeV}$. We show the five contributions to the cross section as defined in Eq.~(\ref{eq:CSDecomp}), as well as their sum.
We show results both for the non-isolated and for the isolated case.}
\end{figure*}

\begin{figure*}
\includegraphics[width=1\textwidth]{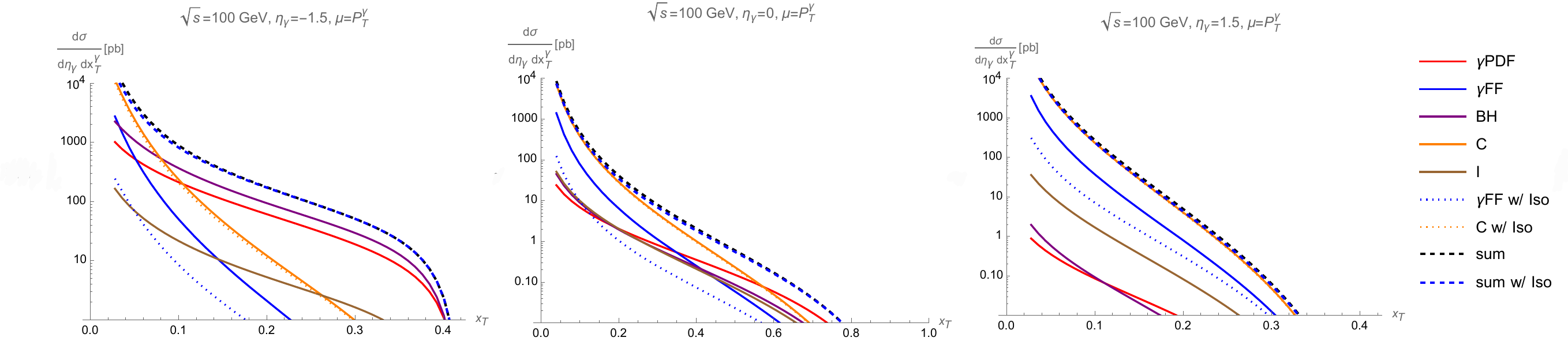}
\caption{\label{fig:CSEICxT} Same as Fig.~\ref{fig:CSEICeta}, but for the dependence on $x_T^\gamma$ for fixed pseudorapidities
$\eta_\gamma = -1.5$ (left), $\eta_\gamma=0$ (center), $\eta_\gamma = 1.5$ (right).}
\end{figure*}

To the best of our knowledge no parameterization is available for the polarized photon-in-proton PDF. In order to obtain an estimate for the longitudinal double-spin asymmetry we adopt a simple model that relates the photon PDF to the gluon one:
\begin{equation}
    g_1^{\gamma/p}(x,\mu) = \fsc\,g_1^{g/p}(x,\mu) \,,\label{eq:Modelg1}
\end{equation}
where $g_1^{g/p}$ denotes the proton's gluon helicity PDF. 
Of course, this can at best be a crude model. Future data may very well shed further light on $g_1^{\gamma/p}$ and falsify this ansatz.
Note that somewhat similar model scenarios for the polarized photon helicity PDF have been used in Refs.~\cite{deFlorian:2023zkc,deFlorian:2024hsu} as input for studies about QED corrections affecting the QCD evolution of parton distributions.

We anticipate a center-of-mass energy of $\sqrt{s}=100\,\mathrm{GeV}$ for electron-proton collisions at the EIC. 
We translate the invariant cross section to 
\begin{equation}
    \frac{\mathrm{d}^2\sigma}{\mathrm{d}\eta_\gamma\,\mathrm{d}x_T^\gamma}=\frac{\pi}{2}\,x_T^\gamma\,s\,E_\gamma\frac{\mathrm{d}^3\sigma}{\mathrm{d}^3\bm{P}_\gamma}\,,\label{eq:EICCS}
\end{equation}
where $\eta_\gamma$ is the photon's pseudorapidity in the c.m.-frame, where we count positive rapidity in the forward proton direction. 
Furthermore, $x_T^\gamma=2p_T^\gamma/\sqrt{s}$ with the 
photon's transverse momentum $p_T^\gamma$. Our plots will show results always for $\mathrm{d}^2\sigma/\mathrm{d}\eta_\gamma\,\mathrm{d}x_T^\gamma$.

\subsection{Unpolarized cross section\label{ssub:EICUnpolCS}}

The dependence of the differential unpolarized cross section (\ref{eq:EICCS}) on the photon's pseudorapidity $\eta_\gamma$ is shown in Fig.~\ref{fig:CSEICeta} along with the various individual contributions of the channels defined in Eq.~(\ref{eq:CSDecomp}). We observe that the 
contributions $\sigma^{\gamma\mathrm{PDF}}$ and $\sigma^{\mathrm{BH}}$ tend to dominate in the backward region (electron direction). 
Hence, this region offers the best opportunity to experimentally learn about the photon-in-proton PDF $f_1^{\gamma/p}(x)$ and to further constrain this function. 
On the other hand, in the forward region (proton direction) $\sigma^{\gamma\mathrm{FF}}$ dominates, along with the Compton channel, 
possibly allowing access to the quark-to-photon fragmentation functions $D_1^{\gamma/q}(z)$. This appears to be the case especially at 
lower $p_T^\gamma$. We further observe that the interference contribution $\sigma^{\mathrm{I}}$ is overall small.

We also study in Fig.~\ref{fig:CSEICeta} the effect of photon isolation on the cross section at the EIC. Here we assume a cone radius of $R=0.7$ 
and an isolation parameter $\xi=0.1$. It is interesting to see that photon isolation indeed strongly suppresses the photon fragmentation channel,
which improves the overall sensitivity to the photon PDF $f_1^{\gamma/p}$. On the other hand, the overall results do not strongly depend on the choice of $R$ as long as $R$ is not too small. We checked explicitly for $R=0.4$ that the plots in Figs.~\ref{fig:CSEICeta}, \ref{fig:CSEICxT} are (almost) unaffected.

The dependence of the unpolarized cross section on the transverse photon momentum is shown in Fig.~\ref{fig:CSEICxT} in the backward, central, 
and forward regions. Again we observe a clear hierarchy of the various contributions, in particular in the backward and forward regions. The photon PDF contribution 
$\sigma^{\gamma\mathrm{PDF}}$ is dominant in the backward region over a large range of transverse photon momenta, in particular for the isolated case. 
Hence, the backward region may be the preferred region for further constraining $f_1^{\gamma/p}$, and the transverse photon momentum dependence appears well-suited for that purpose.

By contrast, the Compton channel and photon fragmentation contributions dominate the transverse photon momentum dependence in the forward region. 
In the mid-rapidity region the hierarchy of channels is not that strict, and the various channels are approximately of similar size for larger transverse photon momenta. However, for smaller transverse momenta of about 2.5 to 10 GeV, the Compton channel and the photon fragmentation contribution dominate as well.

\subsection{Polarized Cross Section\label{ssub:EICPolCS}}

Next we discuss our numerical estimates for the polarized cross section (\ref{eq:LCSDecomp}) at the EIC and the sizes of the various contributions. 
We focus in particular on the sensitivity to the polarized photon-in-proton PDF $g_1^{\gamma/p}$ via the channel  $\Delta \sigma^{\gamma\mathrm{PDF}}$. 
We remind the reader that our results are based on our ad-hoc model ansatz~(\ref{eq:Modelg1}). For the polarized case we will only consider
the isolated cross section, in order to suppress the photon fragmentation contribution. Clearly, for the purpose of studying quark-to-photon fragmentation 
the unpolarized cross section is the preferred observable.

\begin{figure*}
\includegraphics[width=1\textwidth]{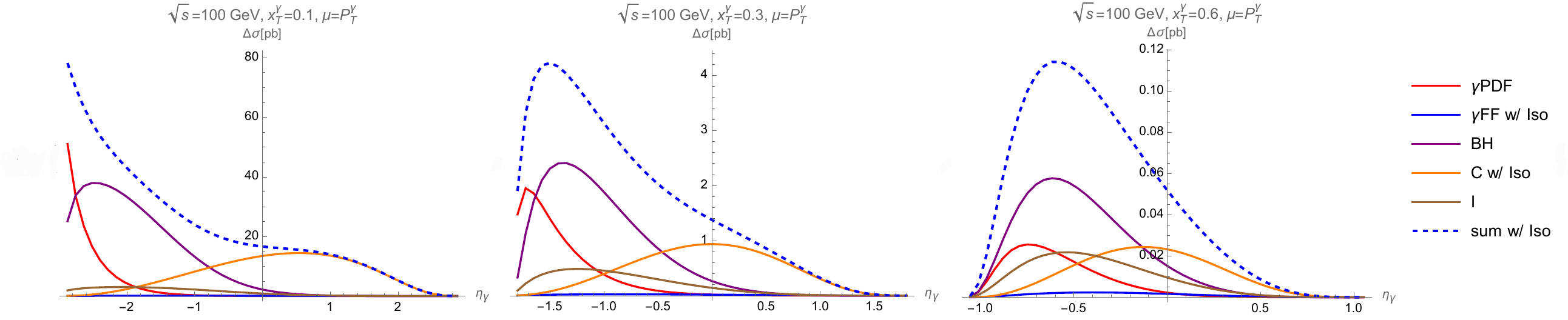}
\caption{\label{fig:LCSEICeta} Same as Fig.~\ref{fig:CSEICeta}, but for the polarized case. We only consider the isolated cross section here.}
\end{figure*}

\begin{figure*}
\includegraphics[width=1\textwidth]{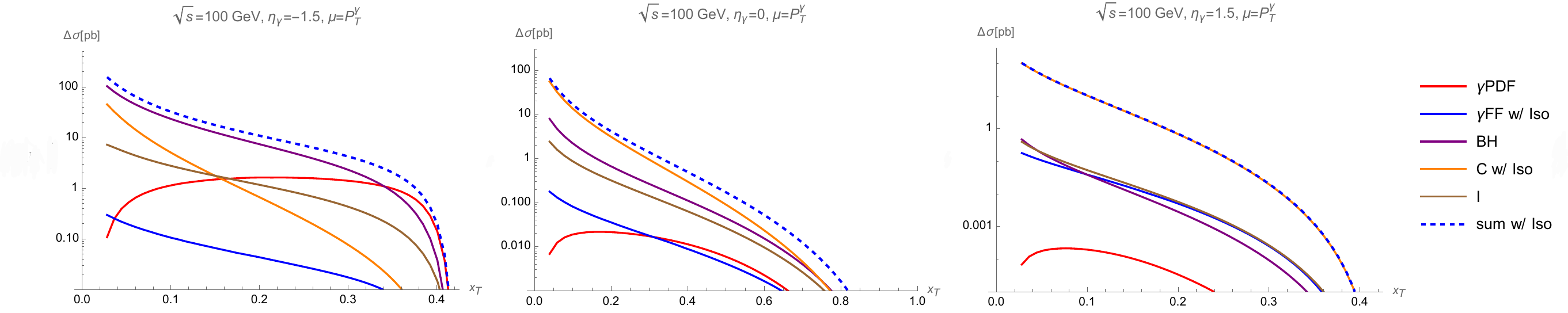}
\caption{\label{fig:LCSEICxT} Same as Fig.~\ref{fig:CSEICxT}, but for the polarized case. We only consider the isolated cross section here.}
\end{figure*}

\begin{figure*}
\includegraphics[width=1\textwidth]{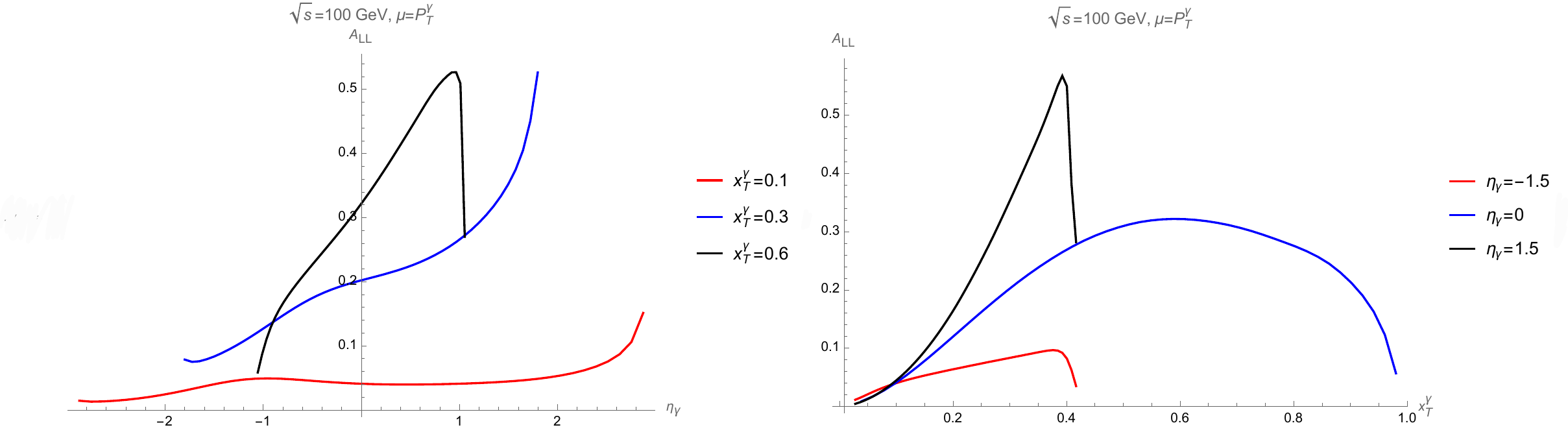}
\caption{\label{fig:ALLEIC} The longitudinal double-spin asymmetry $A_{{\textrm{LL}}}$ at the EIC for c.m.-energy of $\sqrt{s}=100\,\mathrm{GeV}$, 
plotted as function of the photon's pseudorapidity $\eta_\gamma$ for fixed transverse momenta (left) and as function of $x_T^\gamma$ for fixed values of 
pseudorapidity (right).}
\end{figure*}

Figure~\ref{fig:LCSEICeta} shows the dependence of the polarized cross section on the photon's pseudorapidity $\eta_\gamma$, again for three different transverse momenta. The largest contributions of the polarized photon-in-proton PDF to the polarized cross section can be found in the backward region for small to medium transverse photon momenta. The large transverse momentum region is not suitable to constrain $g_1^{\gamma/p}$.
In Fig.~\ref{fig:LCSEICxT} we also show the $p_T^\gamma$ distributions 
for the various individual channels. Again, we see that the backward region is the preferred region for obtaining information on $g_1^{\gamma/p}$.

In experiment, one typically does not directly access the polarized cross section, but rather measures the longitudinal double-spin asymmetry 
$A_{{\textrm{LL}}}$, defined by
\begin{equation}
    A_{{\textrm{LL}}} = \frac{\Delta \sigma}{\sigma}\,.\label{eq:DefALL}
\end{equation}
Figure~\ref{fig:ALLEIC} shows the dependence of the double-spin asymmetry $A_{{\textrm{LL}}}$ on the photon's pseudorapidity $\eta_\gamma$ as well as on the photon's transverse momentum. Overall, large asymmetries of about 20\% to 40\% could be possible. However, in the main region of interest, i.e., the backward region at moderate  transverse momenta, which shows the best sensitivity to the polarized photon-in-proton PDF $g_1^{\gamma/p}$, the size of the asymmetry $A_{{\textrm{LL}}}$ is only about a few percent. However, this might still be sufficient to obtain valuable information on $g_1^{\gamma/p}$.

\section{\label{sec:Conc}Conclusions}

We have analyzed the single-inclusive production of prompt photons in electron-proton scattering at the EIC, computing the spin-averaged as well as
the longitudinally polarized cross section of this reaction to NLO in QCD. At NLO, there are several production channels 
that contribute to both the cross sections.  We have presented analytical results for the partonic cross sections for 
each of the channels. We have also discussed the implementation of a photon isolation cut and its effect on the cross sections.

The purpose of our computation was to investigate the accessibility of the proton's photonic parton distributions at the EIC, the unpolarized photon-in-proton 
PDF $f_1^{\gamma/p}$ and its polarized counterpart $g_1^{\gamma/p}$. In particular, $g_1^{\gamma/p}$ is only poorly known, but could potentially provide (albeit likely small) contributions to the proton spin.

Our numerical studies for the Electron-Ion Collider with a center-of-mass energy of $\sqrt{s}=100\,\mathrm{GeV}$ suggest that the future data 
should very well help to constrain the photonic functions mentioned above, and possibly even the quark-to-photon fragmentation functions, 
provided a careful selection of kinematical regions. 

Future extensions of our work will address the transverse single-spin asymmetry for the process $ep\to \gamma X$, which could be of 
great interest for understanding power-suppressed phenomena in QCD and for accessing the proton's quark and gluon structure at twist-3. 
 
\begin{acknowledgments}

We are grateful to Gunar Schnell for helpful discussions. This work has been supported by Deutsche Forschungsgemeinschaft (DFG) through the Research Unit FOR 2926 (project 409651613).

\end{acknowledgments}

\appendix

\section{\label{app:UCS}Unpolarized partonic cross sections}

Here we list the unpolarized partonic hard-scattering functions that appear in Eqs.~(\ref{eq:resCSgammaPDF})--(\ref{eq:resCSI}).
For the LO terms we have
\begin{eqnarray}
\hat{\sigma}^{\gamma\mathrm{PDF}}(v) &=& \frac{1+v^2}{v}\,,\label{eq:respartCSgammaPDF}\\
\hspace{-0.3cm}\hat{\sigma}^{\gamma\mathrm{FF}}(v,w) &=& \frac{vw\left[1+v^2-2v(1-w)(1+vw)\right]}{(1-v)^2(1-v+vw)}\,.\label{eq:respartCSgammaFF}
\end{eqnarray}
We note (see subsection~\ref{sub:LOandNLO}) that the photon fragmentation channel receives ${\cal O}(\alpha_s)$ corrections that are known from 
Refs.~\cite{Hinderer:2015hra,Hinderer:2017ntk} and will not be recalled here. They are, however, included in our numerical results.

For the Bethe-Heitler contribution we find:
\begin{eqnarray}
\hat{\sigma}^{\mathrm{BH}}\left(v,w,\tfrac{-u}{m_e^2},\tfrac{-u}{\mu^2}\right) & = & C_1^{\textrm{BH}}(v,w)\,\ln\left(\tfrac{-u}{\mu^2}\right)\nonumber\\
&& \hspace{-3cm}+C_2^{\textrm{BH}}(v,w)\,\ln\left(\tfrac{-u}{m_e^2}\right)+C_3^{\textrm{BH}}(v,w)\,\ln(1-v+vw)\nonumber\\
&& \hspace{-3cm}+C_4^{\textrm{BH}}(v,w)\,\ln(1-w)+C_5^{\textrm{BH}}(v,w)\,\ln(w)\nonumber\\
&&\hspace{-3cm}+C_6^{\textrm{BH}}(v,w)\,,\label{eq:partresCSBH}
\end{eqnarray}
with the coefficient functions
\begin{eqnarray}
C_1^{\textrm{BH}}(v,w) & = & \frac{1+v^2}{v}\frac{1+(1-w)^2}{w}\,,\label{eq:C1BH}\nonumber\\
C_2^{\textrm{BH}}(v,w) & = & \frac{1+v^2(1-w)^2}{vw(1-v+vw)^2}\nonumber\\
&&\times(2+2v^2-v(2-w)(2+vw))\,,\label{eq:C2BH}\nonumber\\
C_3^{\textrm{BH}}(v,w) & = & -\frac{2}{vw}(2+v^2(1+(1-w)^2))\,,\label{eq:C3BH}\nonumber\\
C_4^{\textrm{BH}}(v,w) & = &\frac{4(1+v^2)-w(2-w)(1+2v^2)}{vw}\nonumber\\
&&+\frac{2v^2(1-w)-2}{(1-v+vw)^2}\,,\label{eq:C4BH}\nonumber\\
C_5^{\textrm{BH}}(v,w) & = &\frac{2 (1-v^2 (1-w))}{(1-v+vw)^2}+\frac{2-w}{v}\,,\label{eq:C5BH}\nonumber\\
C_6^{\textrm{BH}}(v,w) & = & \frac{2(1-v)(1+2v(1-w))}{(1-v+vw)^2}-4\frac{1-w}{w}\nonumber\\
&&+\frac{1-v}{v}(1-w)+vw\,,\label{eq:C6BH}
\end{eqnarray}

In the Compton channel, we have
\begin{eqnarray}
\hat{\sigma}^{\mathrm{C}}\left(v,w,\tfrac{-u}{m_e^2},\tfrac{-u}{\mu^2}\right) & = & C_1^{\textrm{C}}(v,w)\,\ln\left(\tfrac{-u}{\mu^2}\right)\nonumber\\
&& \hspace{-3.1cm}+C_2^{\textrm{C}}(v,w)\,\ln\left(\tfrac{-u}{m_e^2}\right)+C_3^{\textrm{C}}(v,w)\,\ln\left(\tfrac{1-v}{1-vw}\right)\nonumber\\
&& \hspace{-3.1cm}\,+C_4^{\textrm{C}}(v,w)\,\ln(1-v+vw)\nonumber\\
&&\hspace{-3.1cm}+C_5^{\textrm{C}}(v,w)\,\ln(\tfrac{1-w}{w})+C_6^C(v,w),\label{eq:partresCSC}
\end{eqnarray}
with
\begin{eqnarray}
C_1^{\textrm{C}}(v,w) & = & \frac{v w \left(1+v^2 (1-w)^2\right) }{(1-v)^2 (1-v+vw )^2}\times\nonumber\\
&&(1+ v^2-2v(1-w) (1+ vw ) )\,,\label{eq:C1C}\nonumber\\
C_2^{\textrm{C}}(v,w) & = & \frac{v w (2-v w (2-v w)) }{(1-v)^2 (1-v w)^2}\times\nonumber\\
&&(1+v^2-2vw (1+ v -vw) )\,,\label{eq:C2C}\nonumber\\ 
C_3^{\textrm{C}}(v,w) & = & \frac{2 v w (1+v -2vw)}{(1-v) (1-v w)^2}\,,\label{eq:C3C}\nonumber\\
C_4^{\textrm{C}}(v,w) & = &-\frac{4vw(1-v(1-w)(1-vw))}{(1-v)(1-v+vw)^2}\,,\label{eq:C4C}\nonumber\\
C_5^{\textrm{C}}(v,w) & = & \frac{4vw(2-vw-v^2w+v^2w^2)}{(1-v)^2}\nonumber\\
&&-\frac{2(2-v+v^2 w)}{1-v}+\frac{8-2 v}{1-v+vw}\nonumber\\
&&-\frac{3-v}{(1-v) (1-v w)}-\frac{2 (1-v)}{(1-v+vw)^2}\nonumber\\
&&+\frac{1}{(1-v w)^2}\,,\label{eq:C5C}\nonumber\\
C_6^{\textrm{C}}(v,w) &=& \frac{-vw(2+3v-v^2-4vw)}{(1-v)^2}-\frac{v}{1-v}\nonumber\\
&&-\frac{2 (4-v)}{1-v+vw}+\frac{3 (2-v)}{(1-v) (1-v w)}\nonumber\\
&&+\frac{4(1-v)}{(1-v+vw)^2}-\frac{2}{(1-v
   w)^2}\,,\label{eq:C6C}
\end{eqnarray}

Finally, our results for the interference contribution are
\begin{eqnarray}
\hat{\sigma}^{\textrm{I}}(v,w) &=& C_1^{\textrm{I}}(v,w)\,\ln(1-v)+C_2^{\textrm{I}}(v,w)\,\ln(v)\nonumber\\
&&+C_3^{\textrm{I}}(v,w)\,\ln(1-w)+C_4^{\textrm{I}}(v,w)\,\ln(w)\nonumber\\
&&+C_5^{\textrm{I}}(v,w)\,\ln((1-v w)(1-v+vw)^2)\nonumber\\
&&+C_6^{\textrm{I}}(v,w)\,,\label{eq:partresCSI}
\end{eqnarray}
where
\begin{eqnarray}
C_1^{\textrm{I}}(v,w) & = & \frac{8}{1-v+vw}-4\,,\label{eq:C1I}\nonumber\\
C_2^{\textrm{I}}(v,w) & = & -\frac{4 v^2 (w (3 w-4)+2)-8 v w+8}{(1-v) (1-v+vw)},\label{eq:C2I}\nonumber\\
C_3^{\textrm{I}}(v,w) & = &-\frac{2 v^2 (w (5 w-8)+5)-4 v w+6}{(1-v) (1-v+vw)},\label{eq:C3I}\nonumber\\
C_4^{\textrm{I}}(v,w) & = & \frac{4 v w (1+v-vw)}{(1-v) (1-v+vw)}\,,\label{eq:C4I}\nonumber\\
C_5^{\textrm{I}}(v,w) & = & -\frac{2(1+v-vw)}{1-v}\,,\label{eq:C5I}\nonumber\\
C_6^{\textrm{I}}(v,w) & = & -\frac{2 w (1+v-vw) }{(1-v+vw)^2 (1-v w)}\times\nonumber\\
&&\left(1+v^2-v \left(4-2w-v w^2\right)\right)\,.\label{eq:C6I}
\end{eqnarray}

\section{\label{app:LCS}Longitudinally polarized partonic cross sections}

Here we list the polarized partonic hard cross sections that appear in Eqs.~(\ref{eq:resLCSgammaPDF})--(\ref{eq:resLCSI}). To LO,
\begin{eqnarray}
\Delta\hat{\sigma}^{\gamma\mathrm{PDF}}(v) &=& \frac{1-v^2}{v}\,,\label{eq:respartLCSgammaPDF}\\
\hspace{-0.0cm}\Delta\hat{\sigma}^{\gamma\mathrm{FF}}(v,w) &=& \frac{ v w (1-v+2vw)}{(1-v) (1-v+vw)}\,,\label{eq:respartLCSgammaFF} 
\end{eqnarray}
again with NLO corrections in the fragmentation channel that are known~\cite{Hinderer:2015hra,Hinderer:2017ntk}. 

For the Bethe-Heitler process,
\begin{eqnarray}
\Delta\hat{\sigma}^{\mathrm{BH}}\left(v,w,\tfrac{-u}{m_e^2},\tfrac{-u}{\mu^2}\right) & = & D_1^{\textrm{BH}}(v,w)\,\ln\left(\tfrac{-u}{\mu^2}\right)\nonumber\\
&& \hspace{-3cm}+D_2^{\textrm{BH}}(v,w)\,\ln\left(\tfrac{-u}{m_e^2}\right)+D_3^{\textrm{BH}}(v,w)\,\ln(1-v+vw)\nonumber\\
&& \hspace{-3cm}+D_4^{\textrm{BH}}(v,w)\,\ln(1-w)+D_5^{\textrm{BH}}(v,w)\,\ln(w)\nonumber\\
&&\hspace{-3cm}+D_6^{\textrm{BH}}(v,w)\,,\label{eq:partresLCSBH}
\end{eqnarray}
with the coefficient functions
\begin{eqnarray}
D_1^{\textrm{BH}}(v,w) & = & \frac{1-v^2}{v}(2-w)\,,\label{eq:D1BH}\nonumber\\
D_2^{\textrm{BH}}(v,w) & = & \frac{(2-2v+vw) \left(1+v^2 (1-w)^2\right)}{(1-v+vw)^2}\,,\label{eq:D2BH}\nonumber\\
D_3^{\textrm{BH}}(v,w) & = & 2 v (2-w)\,,\label{eq:D3BH}\nonumber\\
D_4^{\textrm{BH}}(v,w) & = &\frac{2(1-v)}{(1-v+vw)^2}-2v(2-w)\nonumber\\
&&+\frac{2 v}{1-v+vw}+\frac{2-w}{v}\,,\label{eq:D4BH}\nonumber\\
D_5^{\textrm{BH}}(v,w) & = &-\frac{2-2 v^2 (1-w)}{(1-v+vw)^2}-\frac{2-w}{v}\,,\label{eq:D5BH}\nonumber\\
D_6^{\textrm{BH}}(v,w) & = & \frac{4(1-v)}{1-v+vw}+1-w-\frac{3-4w}{v}\nonumber\\
&&-\frac{6 (1-v)}{(1-v+vw)^2}+2 v(1-w)\,,\label{eq:D6BH} 
\end{eqnarray}

The results for the Compton channel are
\begin{eqnarray}
\Delta\hat{\sigma}^{\mathrm{C}}\left(v,w,\tfrac{-u}{m_e^2},\tfrac{-u}{\mu^2}\right) & = & D_1^{\textrm{C}}(v,w)\,\ln\left(\tfrac{-u}{\mu^2}\right)\nonumber\\
&& \hspace{-3.1cm}+D_2^{\textrm{C}}(v,w)\,\ln\left(\tfrac{-u}{m_e^2}\right)+D_3^{\textrm{C}}(v,w)\,\ln(\tfrac{1-v}{1-vw})\nonumber\\
&& \hspace{-3.1cm}+D_4^{\textrm{C}}(v,w)\,\ln(1-v+vw)\nonumber\\
&&\hspace{-3.1cm}+D_5^{\textrm{C}}(v,w)\,\ln(\tfrac{1-w}{w})+D_6^C(v,w),\label{eq:partresLCSC}
\end{eqnarray}
with
\begin{eqnarray}
\hspace{-0.3cm}D_1^{\textrm{C}}(v,w) & = & \frac{v w \left(1+v^2 (1-w)^2\right) (1-v + 2vw)}{(1-v) (1-v+vw)^2}\,,\label{eq:D1C}\nonumber\\
D_2^{\textrm{C}}(v,w) & = & \frac{v^2 w^2 (2-v w) (1+v-2vw)}{(1-v) (1-v w)^2}\,,\label{eq:D2C}\nonumber\\
D_3^{\textrm{C}}(v,w) & = & -\frac{2 v w (1+v-2 v w)}{(1-v) (1-v w)^2}\,,\label{eq:D3C}\nonumber\\
D_4^{\textrm{C}}(v,w) & = & \frac{4vw(1-v(1-w)(1-vw))}{(1-v)(1-v+vw)^2}\,,\label{eq:D4C}\nonumber\\
D_5^{\textrm{C}}(v,w) & = & \frac{v w }{(1-v) (1-v+vw)^2 (1-v w)^2} \times\nonumber\\
&&\left[1-v+2vw +v^2(1-2w-3w^2) \right.\nonumber\\
&&\left.-2 v^4 w (1-w) (2 w-1)\times\right.\nonumber\\
&&\left.(2-w-vw(1-w))\right.\nonumber\\
&&\left.+v^3 \left((7-4 w) w^2-1\right)\right]\,,\label{eq:D5C}\nonumber\\
D_6^{\textrm{C}}(v,w) &=& \frac{8-2 v}{1-v+vw}+\frac{v(1+2w+vw)}{1-v}\nonumber\\
&&+\frac{5 v-8}{(1-v) (1-v w)}+\frac{4 (v-1)}{(1-v+vw)^2}\nonumber\\
&&+\frac{4}{(1-v w)^2}\,,\label{eq:D6C} 
\end{eqnarray}

The interference contribution reads
\begin{eqnarray}
\Delta\hat{\sigma}^{\textrm{I}}(v,w) &=& D_1^{\textrm{I}}(v,w)\,\ln(1-v)+D_2^{\textrm{I}}(v,w)\,\ln(v)\nonumber\\
&&+D_3^{\textrm{I}}(v,w)\,\ln(1-w)+D_4^{\textrm{I}}(v,w)\,\ln(w)\nonumber\\
&&+D_5^{\textrm{I}}(v,w)\,\ln((1-v w)(1-v+vw)^2)\nonumber\\
&&+D_6^{\textrm{I}}(v,w)\,,\label{eq:partresLCSI}
\end{eqnarray}
with the coefficients
\begin{eqnarray}
D_1^{\textrm{I}}(v,w) & = & 4-\frac{8}{1-v+vw}\,,\label{eq:D1I}\nonumber\\
D_2^{\textrm{I}}(v,w) & = & -\frac{4 v w (2-v w)}{(1-v) (1-v+vw)}\,,\label{eq:D2I}\nonumber\\ 
D_3^{\textrm{I}}(v,w) & = &\frac{-2-4vw+2 v^2 \left(1+w^2\right)}{(1-v) (1-v+vw)}\,,\label{eq:D3I}\nonumber\\ 
D_4^{\textrm{I}}(v,w) & = & -\frac{4 v w (1+v-vw)}{(1-v) (1-v+vw)}\,,\label{eq:D4I}\nonumber\\
D_5^{\textrm{I}}(v,w) & = & \frac{2+2 v (1-w)}{1-v}\,,\label{eq:D5I}\nonumber\\
D_6^{\textrm{I}}(v,w) & = & \frac{2 w (1+v-vw) }{(1-v+vw)^2 (1-v w)}\nonumber\nonumber\\
&&\left[1-4v+v^2+v w \left(2+v w\right)\right]\,.\label{eq:D6I}
\end{eqnarray}

\bibliography{Referenzen}

\end{document}